\documentclass[a4paper,fleqn,usenatbib,useAMS]{mnras}

\usepackage[T1]{fontenc}
\usepackage{ae,aecompl}

\usepackage{graphicx}	
\usepackage{amsmath}	
\usepackage{amssymb}	
\usepackage{footnote}
\usepackage{caption}
\usepackage{color}
\usepackage{bm} 

\def\OI{[\mbox{O\,{\sc i}}]~$\lambda 6300$}
\def\OIII{[\mbox{O\,{\sc iii}}]~$\lambda 5007$}

\def\NII{[\mbox{N\,{\sc ii}}]~$\lambda 6583$}

\def\OI{[\mbox{O{\sc i}}]~$\lambda 6300$}

\def\SIIa{[\mbox{S{\sc ii}}]~$\lambda 6716$}
\def\SIIb{[\mbox{S{\sc ii}}]~$\lambda 6731$}

\def\Ha{{H$\alpha$}}
\def\Hb{{H$\beta$}}

\def\NIIHa{[\mbox{N\,{\sc ii}}]~$\lambda 6583$/H$\alpha$}
\def\SIIHa{[\mbox{S\,{\sc ii}}]~$\lambda\lambda 6716,6731$/H$\alpha$}
\def\NeIIIHb{[\mbox{Ne\,{\sc iii}}]~$\lambda 3869$/H$\beta$}
\def\HeIIHb{\mbox{He\,{\sc ii}}~$\lambda 4686$/H$\beta$}

\def\OIHa{[\mbox{O\,{\sc i}}]~$\lambda 6300$/H$\alpha$}
\def\OIIIHb{[\mbox{O\,{\sc iii}}]~$\lambda 5007$/H$\beta$}

\def\Edd{$\lambda_{\rm Edd}$}
\def\Mbh{$M_{\rm BH}$}

\def\LOIII{$L[\mbox{O\,{\sc iii}}]$}

\def\LOIIIs4{$L[\mbox{O\,{\sc iii}}]$/$\sigma^4$}

\def\Mbh{$M_{\rm BH}$}

\def\kms{${\rm km}~{\rm s}^{-1}$}
\newcommand{\ergcms}	{\ifmmode {\rm erg\,cm}^{-2}\,{\rm s}^{-1} \else erg\,cm$^{-2}$\,s$^{-1}$\fi}

\newcommand{  \Ntot    }{642} 
\newcommand{  \Nmbh    }{473}
\newcommand{  \Nmbhper    }{74\%}

\definecolor{myblue}{RGB}{0, 100, 200}
\definecolor{myblue2}{RGB}{0, 100, 220}

\title[BASS-III: Emission lines vs. Eddington ratio]{BAT AGN Spectroscopic Survey-III. An observed link between AGN Eddington ratio and narrow emission line ratios}

\author[K. Oh et al.]{
Kyuseok Oh$^{1}$\thanks{E-mail: ohk@phys.ethz.ch},
Kevin Schawinski$^{1}$,
Michael Koss$^{1,11}$,
Benny Trakhtenbrot$^{1,12}$,
\newauthor Isabella Lamperti$^{1}$,
Claudio Ricci$^{2}$,
Richard Mushotzky$^{3}$,
Sylvain Veilleux$^{3}$,
\newauthor Simon Berney$^{1}$,
D. Michael Crenshaw$^{4}$,
Neil Gehrels$^{5}$,
Fiona Harrison$^{6}$,
\newauthor Nicola Masetti$^{7,8}$,
Kurt T. Soto$^{1}$,
Daniel Stern$^{9}$,
Ezequiel Treister$^{2}$,
Yoshihiro Ueda$^{10}$
\\
$^{1}$Institute for Astronomy, Department of Physics, ETH Zurich, Wolfgang-Pauli-Strasse 27, CH-8093 Zurich, Switzerland \\
$^{2}$Instituto de Astrof\'{\i}sica, Facultad de F\'{\i}sica, Pontificia Universidad Cat\'olica de Chile, Casilla 306, Santiago 22, Chile\\
$^{3}$Astronomy Department and Joint Space-Science Insitute, University of Maryland, College Park, MD, USA\\
$^{4}$Department of Physics and Astronomy, Georgia State University, Astronomy Offices, One Park Place South SE, Suite 700, Atlanta, GA 30303, USA\\
$^{5}$NASA Goddard Space Flight Center, Greenbelt, MD 20771, USA\\
$^{6}$Cahill Center for Astronomy and Astrophysics, California Institute of Technology, Pasadena, CA 91125, USA\\
$^{7}$INAF - Istituto di Astrofisica Spaziale e Fisica Cosmica di Bologna, via Gobetti 101, 40129 Bologna, Italy\\
$^{8}$Departamento de Ciencias F\'{\i}sicas, Universidad Andr\'es Bello, Fern\'andez Concha 700, Las Condes, Santiago, Chile\\
$^{9}$Jet Propulsion Laboratory, California Institute of Technology, 4800 Oak Grove Drive, MS 169-224, Pasadena, CA 91109, USA\\
$^{10}$Department of Astronomy, Kyoto University, Kyoto 606-8502, Japan\\
$^{11}$Ambizione fellow\\
$^{12}$Zwicky fellow
}
\date{Accepted XXX. Received YYY; in original form ZZZ}

\pubyear{2016}

\begin{document}
\label{firstpage}
\pagerange{\pageref{firstpage}--\pageref{lastpage}}
\maketitle

\begin{abstract}
We investigate the observed relationship between black hole mass ($M_{\rm BH}$), bolometric luminosity ($L_{\rm bol}$), and Eddington ratio (\Edd) with optical emission line ratios (\NIIHa, \SIIHa, \OIHa, \OIIIHb, \NeIIIHb, and \HeIIHb) of hard X-ray-selected AGN from the BAT AGN Spectroscopic Survey (BASS). We show that the \NIIHa\ ratio exhibits a significant correlation with \Edd\ ($R_{\rm Pear}=-0.44$, $p$-value=$3\times10^{-13}$, $\sigma=0.28$ dex), and the correlation is not solely driven by $M_{\rm BH}$ or $L_{\rm bol}$. The observed correlation between \NIIHa\ ratio and $M_{\rm BH}$ is stronger than the correlation with $L_{\rm bol}$, but both are weaker than the $\lambda_{\rm Edd}$ correlation. This implies that the large-scale narrow lines of AGN host galaxies carry information about the accretion state of the AGN central engine. We propose that the \NIIHa\ is a useful indicator of Eddington ratio with 0.6 dex of rms scatter, and that it can be used to measure $\lambda_{\rm Edd}$ and thus $M_{\rm BH}$ from the measured $L_{\rm bol}$, even for high redshift obscured AGN. We briefly discuss possible physical mechanisms behind this correlation, such as the mass-metallicity relation, X-ray heating, and radiatively driven outflows. 

\end{abstract}




\section{Introduction}



Nebular emission lines are a powerful tool for diagnosing the physical state of ionized gas and studying central nuclear activity. Optical emission line ratios can be used to discriminate between emission from the star formation in galaxies and harder radiation such as from the central nuclear activity around a supermassive black holes \citep[e.g.,][]{Baldwin81,Veilleux87,Kewley01, Kauffmann03}. Compared to star forming galaxies, active galactic nuclei (AGN) produce greater numbers of higher energy photons (e.g., UV and X-rays) and, therefore drive higher ratios of the collisionally excited forbidden lines compared to the photoionization-induced Balmer emission lines. Although such line ratios provide useful AGN diagnostics, even for obscured AGN \citep{Reyes08, Yuan16}, they may not be effective in selecting all heavily obscured AGN and/or AGN that lack significant amounts of low density gas \citep{Elvis81, Iwasawa93, Griffiths95,  Barger01, Comastri02, Rigby06, Caccianiga07}.

\begin{figure*}
	\includegraphics[width=\linewidth]{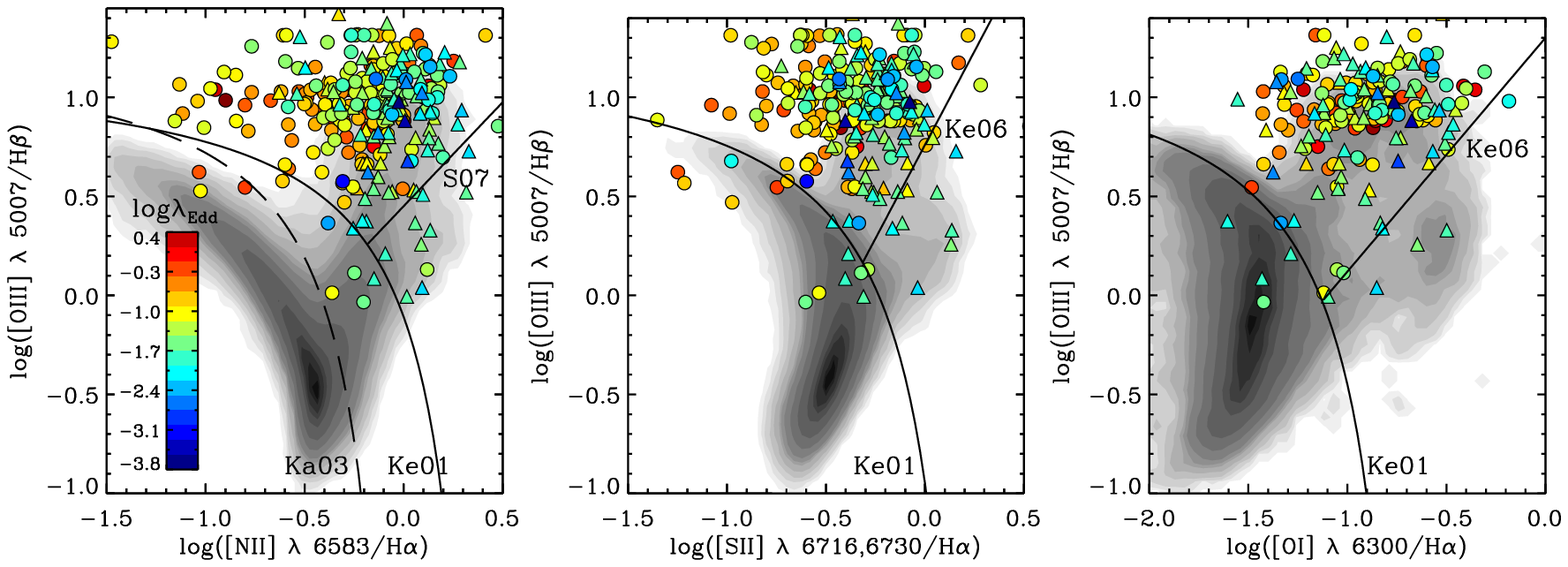} 
    \caption[Caption for LOF]{Emission line diagnostic diagrams for the BASS sources with signal-to-noise (S/N) ratio $>3$. Left: The \NIIHa\ versus \OIIIHb\ diagnostic diagram. 
    Colour filled circles and triangles indicate type 1 AGNs (including type 1.9) and type 2 AGNs, respectively. 
    The empirical star-formation curve obtained from \citet{Kauffmann03} (dashed curve) and the theoretical maximum starburst model of \citet{Kewley01} (solid curve) are used. 
    The solid-straight line is the empirical demarcation of \citet{Schawinski07} distinguishing the Seyfert AGN from the LINERs. 
    The Eddington ratio of BASS sources is shown with color-filled dots.
    Middle: The \SIIHa\ versus \OIIIHb\ diagnostic diagram. 
    Right: The \OIHa\ versus \OIIIHb\ diagnostic diagram. Demarcation lines from \citet{Kewley01, Kewley06} are used. 
    In all panels we also show more than 180,000 SDSS emission-line galaxies with filled contours chosen from the OSSY catalog ($z<0.2$) with S/N $>3$ for \NII, \Ha, \OIII, \Hb, \SIIa, \SIIb, and \OI. 
    }
    \label{fig:bpt_Edd}
\end{figure*}

With the recent advent of hard X-ray ($>10$\,keV) facilities, such as \textit{INTEGRAL} \citep{Winkler03}, \textit{Swift} \citep{Gehrels04} and \textit{NuSTAR} \citep{Harrison13}, it is now possible to study samples of AGN that are less biased to obscuration and include even Compton thick sources ($N_{\rm H}>10^{24} {\rm cm}^{-2}$, \citealt{Ricci15, Koss16}). In particular, the Burst Alert Telescope (BAT, \citealt{Barthelmy05}) on board the \textit{Swift} satellite has been observing the sky in the 14-195 keV energy band since 2005, reaching sensitivities of $1.3\times10^{-11}\ergcms$ over 90\% of the sky. The 70 month \textit{Swift}-BAT all-sky hard X-ray survey\footnote{http://heasarc.gsfc.nasa.gov/docs/swift/results/bs70mon/} detected 1210 objects, of which 836 are AGN \citep{Baumgartner13}. While the BAT detection is relatively unabsorbed up to Compton thick levels (e.g., $N_{\rm H}<10^{24} {\rm cm}^{-2}$, \citealt{Koss16}) heavily Compton thick AGN ($N_{\rm H}>10^{25} {\rm cm}^{-2}$) are missed by X-ray surveys but may sometimes be detected using optical emission line diagnostics and strong \OIII\ emission lines (e.g., \citealt{Maiolino98}).

\begin{figure*}
	\includegraphics[width=\linewidth]{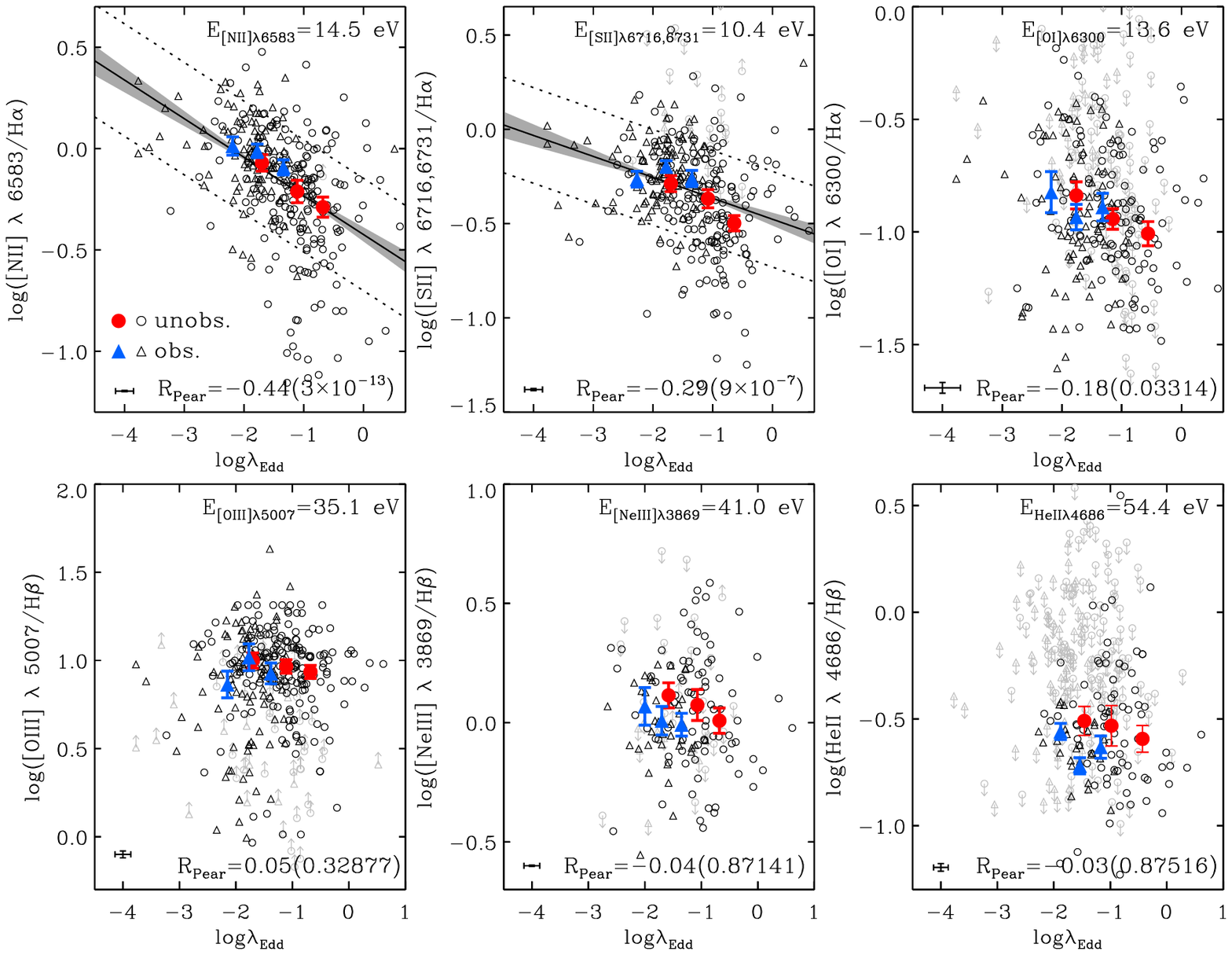}
    \caption{Optical emission line ratio versus Eddington ratio diagram.  
    Black open circles and triangles indicate type 1 AGN (including type 1.9) and type 2 AGN, respectively. 
    Median at each bin is shown with colour-filled symbols. 
    Bin size is determined to have at least 10 sources. 
    Black solid lines indicate the Eddington ratio - optical emission line ratio relations (equation~\ref{eq:regression}). 
    The grey shaded regions account for the errors in the slope and intercept of the relation. 
    The rms deviation is shown with dotted lines. 
    An error bar in the bottom-left corner at each panel indicates typical uncertainties in Eddington ratio and optical emission line ratio. The ionization potential for each emission line is shown in the legends. Also, Pearson correlation coefficients and $p$-values are shown in the bottom-right corner of each panel. 
    An emission line detection at S/N$<3$ (upper- or lower-limit) is shown with grey symbols. 
    }
    \label{fig:ratio_Edd}
\end{figure*}

The relationship between Eddington ratio ($\lambda_{\rm Edd}\equiv L/L_{\rm Edd}$, where $L_{\rm Edd}\equiv 1.3 \times 10^{38} M_{\rm BH}/M_{\odot}$) and the position of AGN in emission-line diagrams is an important topic of study because of the difficulty in measuring black hole mass ($M_{\rm BH}$) from velocity dispersion in high redshift AGN. \citet{Kewley06} investigated host properties of nearby emission-line galaxies ($0.04<z<0.1$) from the SDSS. They found that the $\lambda_{\rm Edd}$ (inferred from $L_{\rm [OIII]}/{\sigma_{\star}}^{4}$, where $\sigma_{\star}$ is a stellar velocity dispersion) shows an increase with $\phi$, a measure of distance from the LINER regime in the \OIIIHb\ vs. \OIHa\ diagram. Similarly, an SDSS study of unobscured AGN by \citet{Stern13} found a dependence of emission-line diagnostics on the $\lambda_{\rm Edd}$. However, the estimation of $\lambda_{\rm Edd}$ and the introduced relationship between the angle $\phi$ and $L_{\rm [OIII]}/{\sigma_{\star}}^{4}$ were both dependent on the strength of \OIII. Also, the previous studies did not take into account X-ray selection focusing on the large sample of optically selected AGN. Both highly ionized optical emission lines and X-rays are thought to be a measure of the AGN bolometric luminosity. However, hard X-rays are less biased against dust obscuration and the contribution from star-forming activity than optical emission lines.

The BAT AGN Spectroscopic Survey (BASS) Data Release 1 (Koss et al., in submitted) compiled 642 optical spectra of nearby AGN ($\langle z \rangle \sim 0.05$) from public surveys (SDSS, 6dF; \citealt{Abazajian09, Jones09, Alam15}) and dedicated follow-up observations (e.g., from telescopes at the Kitt Peak, Gemini, Palomar, and SAAO observatories). The data release provided emission line measurements as well as $M_{\rm BH}$ and $\lambda_{\rm Edd}$ estimates for the majority of obscured and un-obscured AGN (\Nmbhper, \Nmbh/\Ntot), including 340 AGN with $M_{\rm BH}$ measurements reported for the first time.

In this paper, we use the BASS measurements to investigate the observed relationship between black hole mass ($M_{\rm BH}$), bolometric luminosity ($L_{\rm bol}$), and Eddington ratio (\Edd) with optical emission line ratios (\NIIHa, \SIIHa, \OIHa, \OIIIHb, \NeIIIHb, and \HeIIHb) for both obscured and unobscured AGN.  

We assume a cosmology with $h=0.70$, $\Omega_{\rm M}=0.30$, and $\Omega_{\Lambda}=0.70$ throughout this work.

\begin{table*}
\centering
\begin{minipage}{160mm}
	\caption{Bayesian linear regression fit.}
	\label{tab:regression}
	\begin{tabular}{llcccccc} 
		\hline
	line ratio &  N &	$\alpha$ & $\beta$ & RMSD & $R_{\rm Pear}$(\textit{p}-value) & $R_{\rm Pear, unobs}$(\textit{p}-value) & $R_{\rm Pear, obs}$(\textit{p}-value) \\
	(1)		  &	(2)	& (3) & (4)		 & (5)	     & (6)	 & (7) & (8)   \\
	\hline
	\NIIHa		& 297	& $-0.42\pm0.04$		& $-0.19\pm0.02$		& 0.28 	          		& -0.44 ($3 \times10^{-13}$) & -0.34 ($0.00002$) 	& -0.28 ($0.00128$) \\
	\SIIHa 		& 288	& $-0.48\pm0.03$		& $-0.11\pm0.02$		& 0.25 		    		& -0.29 ($9 \times 10^{-7}$) & -0.26 ($0.00080$) 	&  0.11 ($0.56180$) \\
	\OIHa		& 205	& $\cdot \cdot \cdot$	& $\cdot \cdot \cdot$	& $\cdot \cdot \cdot$ 	& (0.03314) 				 & (0.02777)			& (0.36499) \\
	\OIIIHb		& 286	& $\cdot \cdot \cdot$	& $\cdot \cdot \cdot$	& $\cdot \cdot \cdot$	& (0.32877) 				 & (0.38456) 			& (0.34875) \\
	\NeIIIHb 	& 125	& $\cdot \cdot \cdot$	& $\cdot \cdot \cdot$	& $\cdot \cdot \cdot$ 	& (0.87141)					 & (0.38163)			& (0.78629) \\
	\HeIIHb		& 107	& $\cdot \cdot \cdot$	& $\cdot \cdot \cdot$	& $\cdot \cdot \cdot$ 	& (0.87516)		 			 & (0.56490)		 	& (0.08583) \\
	\hline
	\end{tabular}
    \medskip
\\Note. (1) optical emission line ratio; (2) size of sample; (3) intercept; (4) slope; (5) rms deviation; (6) Pearson $R$ coefficient and \textit{p}-value; (7) Pearson $R$ coefficient and \textit{p}-value for unobscured AGN; (8) Pearson $R$ coefficient and \textit{p}-value for obscured AGN.
\end{minipage}
\end{table*}

\begin{figure*}
	\includegraphics[width=1.0\textwidth]{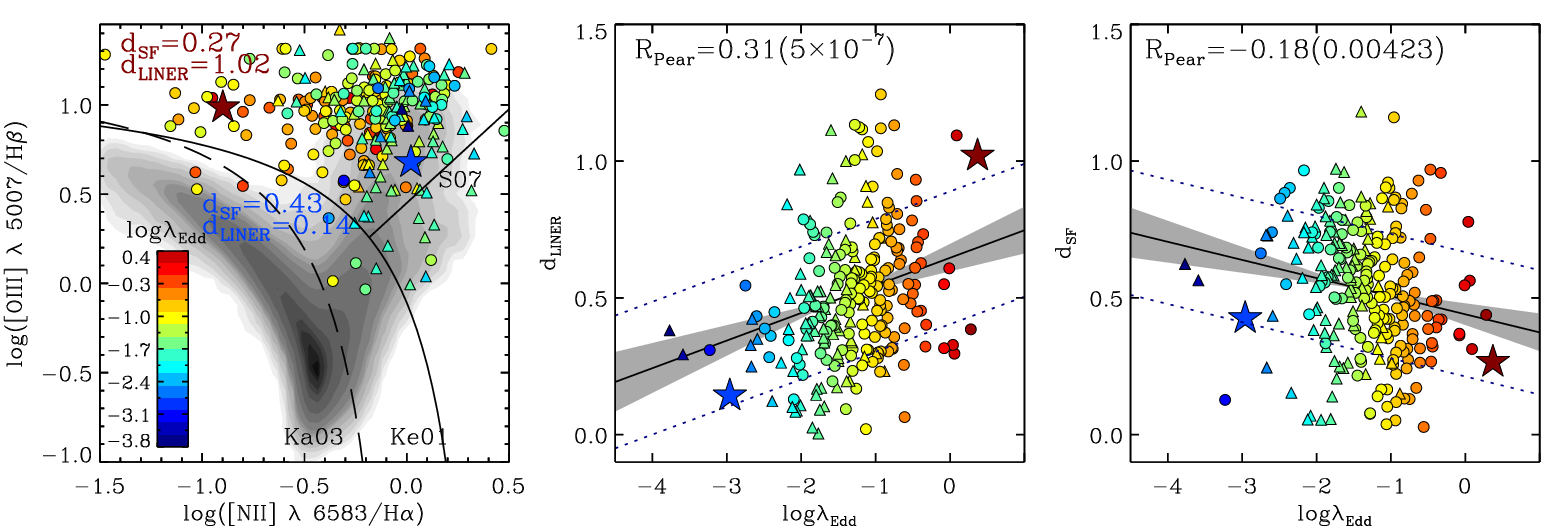}
    \caption{${d}_{\rm LINER}$ and ${d}_{\rm SF}$ as a function of Eddington ratio. Left panel illustrates distances of two examples (star symbols) with corresponding color-coded Eddington ratio as in Fig.~\ref{fig:bpt_Edd}. Middle and right panels show the ${d}_{\rm LINER}$ and ${d}_{\rm SF}$ distributions, respectively. 
    Bayesian linear regression fit, errors in the slope, intercept of the fit, and the rms deviation are shown with black straight lines, grey shaded regions and dotted lines, as in Fig.~\ref{fig:ratio_Edd}.
    }
    \label{fig:distance}
\end{figure*}

\section{Sample Selection, Data, and Measurements}
\label{sec:bass} 

In this section, we briefly summarize the measurement procedures for optical emission lines, $M_{\rm BH}$, and $\lambda_{\rm Edd}$. The BASS DR1 measured nebular emission line strengths by performing a power-law fit with Gaussian components to model the continuum and emission lines. For unobscured AGN, two Gaussian components are allowed in the \Ha\ and \Hb\ emission line regions to account for both broad (${\rm FWHM}>1000$ \kms) and narrow (${\rm FWHM}<1000$ \kms) components. When broad \Hb\ is detected, \Mbh\ is measured using the single-epoch method following \citet{Trakhtenbrot12}. If no broad \Hb\ is detected, \Mbh\ is measured based on the line width and luminoisty of broad \Ha\ (equation 9 from \citealt{Greene05}). For obscured AGN, the estimation of \Mbh\ relies on the close correlations between \Mbh\ and the stellar velocity dispersion ($\sigma_{\star}$, e.g., \citealt{Kormendy13}). Stellar velocity dispersion is derived from the penalised pixel fitting method (\texttt{pPXF}, \citealt{Cappellari04}) by implementing a modified version of the masking procedure introduced for the analysis of SDSS DR7 \citep{Abazajian09} galaxy spectra (the OSSY catalog\footnote{http://gem.yonsei.ac.kr/ossy/}, \citealt{Sarzi06, Oh11, Oh15}). 

Since the obscuration mostly affect the estimation of $L_{\rm bol}$ for Compton thick AGN ($N_{\rm H} > 10^{24} cm^{-2}$), we estimated $L_{\rm bol}$ from the intrinsic (i.e., absorption and k-corrected) 14-150 keV luminosities reported in \citet{Ricci15} and Ricci et al. (in prep.), transforming them into 14-195 keV luminosities assuming a power-law continuum with a photon index of Gamma=1.9. After converting the 14-195 keV luminosity to the intrinsic 2-10 keV luminosity the procedure described by following \citet{Rigby09}, we then applied the median bolometric correction introduced by \citet{Vasudevan09}. It is noteworthy to mention that the estimation of $L_{\rm bol}$ comes solely from hard X-ray band (14-195 keV) and its constant conversion factor ($k=8$). We briefly discuss the effect of different $L_{\rm bol}$ estimation in Section 3. We then combined the measured $M_{\rm BH}$ with the $L_{\rm bol}$ to calculate $\lambda_{\rm Edd}$ ($\lambda_{\rm Edd} \equiv L_{\rm bol}/L_{\rm Edd}$) assuming $L_{\rm Edd} \equiv 1.3 \times 10^{38} (M_{\rm BH}/M_{\odot})$. For more details, refer to the first data release (Koss et al., in submitted).

We focus on the sub-sample of the 642 optical spectra from the BASS DR1. We consider only non-beamed AGN, which were selected by cross-matching the BASS sources with the Roma blazar catalog (BZCAT) v5.0 \citep{Massaro09}.  We then restricted our samples to  redshifts of $0.01<z<0.40$ to have coverage of the \Hb\ and \Ha\ region. \citet{Berney15} investigated the effect of slit size for the BASS DR1 sources and showed that the ratio between extinction corrected \LOIII\ and $L_{\rm 14-195 keV}$ is constant when excluding the nearest galaxies ($z<0.01$) while the scatter slightly decreases towards larger slit sizes. We used the same redshift range following this approach. However, it should be noted that aperture effect does not change our results shown in Section~\ref{section:result}. We tested whether sources with large physical coverage ($>2$ kpc) still found a significant correlation in a smaller sample size suggesting that slit size is not important for this study. We also selected only spectral fits with a good quality as listed in the DR1 tables (Koss et al., in submitted). We note that sources with spectra taken from the 6dF Galaxy Survey \citep{Jones09} are only used to derive emission line ratios and to measure stellar velocity dispersions (e.g., \citealt{Campbell14}) due to the lack of flux calibration as necessary for broad line black hole mass measurements. Samples sizes for each emission line ratio used in this paper are listed in Table~\ref{tab:regression}.

\section{Relations between optical emission line ratios and basic AGN properties}
\label{section:result}
Fig.~\ref{fig:bpt_Edd} shows the emission-line diagnostic diagrams for the BASS sources according to $\lambda_{\rm Edd}$ (colour-coded). The majority of the BASS sources ($>90\%$) are found in the Seyfert region in each panel. 

In order to study the statistical significance of any correlations with $\lambda_{\rm Edd}$, we show optical emission line ratios as a function of $\lambda_{\rm Edd}$ in Fig.~\ref{fig:ratio_Edd}. We performed Bayesian linear regression fit (equation~\ref{eq:regression}) to all points using the method of \citet{Kelly07} which accounts for measurement errors in both axes. The relation between $\lambda_{\rm Edd}$ and optical emission line ratio (black solid line in Fig.~\ref{fig:ratio_Edd}) is determined by taking the median of 10,000 draws from the posterior probability distribution of the converged parameters (intercept and slope). The errors of intercept and slope are reported from 1$\sigma$ confidence ellipse. The root-mean-square (rms) deviation is shown with black dotted lines at each panel. 
\begin{equation}
	  \log(F_{\rm line}/F_{\rm Balmer})= \alpha + \beta\log\lambda_{\rm Edd} 	
\label{eq:regression}
\end{equation} 
The values of $\alpha$ (intercept),  $\beta$ (slope), Pearson correlation coefficient, rms deviation, and \textit{p}-value are summarised in Table~\ref{tab:regression}.

We find that the $\lambda_{\rm Edd}$ is significantly anti-correlated with optical emission line ratios for both the \NIIHa\ and \SIIHa\ ratios but not for the other line ratios. The larger the $\lambda_{\rm Edd}$, the smaller the line ratio of \NIIHa\ and \SIIHa. We find that Pearson $R$ coefficient and $p$-value of the anti-correlation between \NIIHa\ and $\lambda_{\rm Edd}$ are $-0.44$ and $3\times10^{-13}$, respectively, with 0.28 dex of rms deviation. We also found a significantly anti-correlated relationship for both the 
\NIIHa\ and \SIIHa\ ratios with a more stringent S/N cut of optical emission lines ($>10$). AGN variability may induces the scatter shown in the anti-correlation between \NIIHa\ and $\lambda_{\rm Edd}$. Since X-ray emission that we used to derive $L_{\rm bol}$ and $\lambda_{\rm Edd}$ has different time-scales compared to optical narrow emission lines, a scatter around the anti-correlation can be explained \citep{Mushotzky93, Schawinski15}. Also, differences in metallicities and/or structures of the narrow-line regions may contribute to the scatter shown above. In order to quantitatively investigate if the $\lambda_{\rm Edd}$ shows a stronger anti-correlation with \NIIHa\ or with \SIIHa, we run a $z$-test based on the two Pearson correlation coefficients (Fisher $r$-to-$z$ transformation). The $p$-value (0.019) suggests that \NIIHa\ shows a significantly stronger anti-correlation than \SIIHa\ with $\lambda_{\rm Edd}$. 

Moreover, we also find that the observed anti-correlation between \NIIHa\ and $\lambda_{\rm Edd}$ is valid for obscured AGN (blue filled triangles in Fig.~\ref{fig:ratio_Edd}) as well as unobscured AGN (red filled circles in Fig.~\ref{fig:ratio_Edd}). Pearson $R$ coefficient and $p$-value for obscured AGN are $-0.28$ and 0.00128, respectively. For unobscured AGN, we report $-0.34$ and 0.00002 as Pearson $R$ coefficient and $p$-value. We report that $\lambda_{\rm Edd}$ can be estimated from the measured \NIIHa\ ratio as follows, with 0.6 dex of rms deviation: 
\begin{equation}
	\log\lambda_{\rm Edd} = (-1.52\pm0.04) + (-1.00\pm0.13)\times\log([{\mbox{N\,\textsc{ii}}}]\lambda 6583/{\rm H}\alpha)
\label{eq:indicator}
\end{equation} 

Another way to study the location on the emission line diagnostic diagram is to measure the shortest distance from the star-forming and LINER lines in \NIIHa.  The $\lambda_{\rm Edd}$ distribution shown in the \NIIHa\ emission line diagnostic diagram (left panel of Fig.~\ref{fig:bpt_Edd}) enables us to infer that AGN falling in the Seyfert region exhibit different $\lambda_{\rm Edd}$ according to their location, i.e., combinations of emission line ratios. We define the distance between the location of a given object and the empirical demarcation line of \citet{Schawinski07} ($d_{\rm LINER}$) and the theoretical maximum starburst model of \citet{Kewley01} ($d_{\rm SF}$). The separation between Seyfert and LINER was obtained by visual determination based on \NIIHa\ vs. \OIIIHb\ diagram for nearly 50,000 nearby SDSS galaxies ($0.05<z<0.10$, \citealt{Schawinski07}). For Seyfert and LINERs classified using the \SIIHa\ and \OIHa\ diagrams, the authors determined the demarcation line in the \NIIHa\ diagram. In particular, we measured $d_{\rm SF}$ by moving the demarcation line of \citet{Kewley01} in parallel with the original one until it matches the location of the given object. The measured distance, $d_{\rm LINER}$, that originated from optical emission line ratios which depict the physical state of the innermost region of the galaxy is a function of the $\lambda_{\rm Edd}$ (middle panel in Fig.~\ref{fig:distance}). We report Pearson $R$ coefficient and $p$-value for $d_{\rm LINER}$ and ${\rm log}\lambda_{\rm Edd}$ with $0.31$ and $5\times10^{-7}$ while $d_{\rm SF}$ shows less significant statistics ($R_{\rm Pear}= -0.18$, $p$-value=0.00423) suggesting that $\lambda_{\rm Edd}$ is less likely a function of $d_{\rm SF}$. We also ran a test to see if sources with extreme $\lambda_{\rm Edd}$ were driving correlations we found. 
For this test, we used a limited range of $\lambda_{\rm Edd}$ ($-2.67<log\lambda_{\rm Edd}<0.00$) which excludes small number of objects shown at both high- and low-end of $\lambda_{\rm Edd}$ and we found a significant correlation.

\begin{figure*}
	\includegraphics[width=0.79\textwidth]{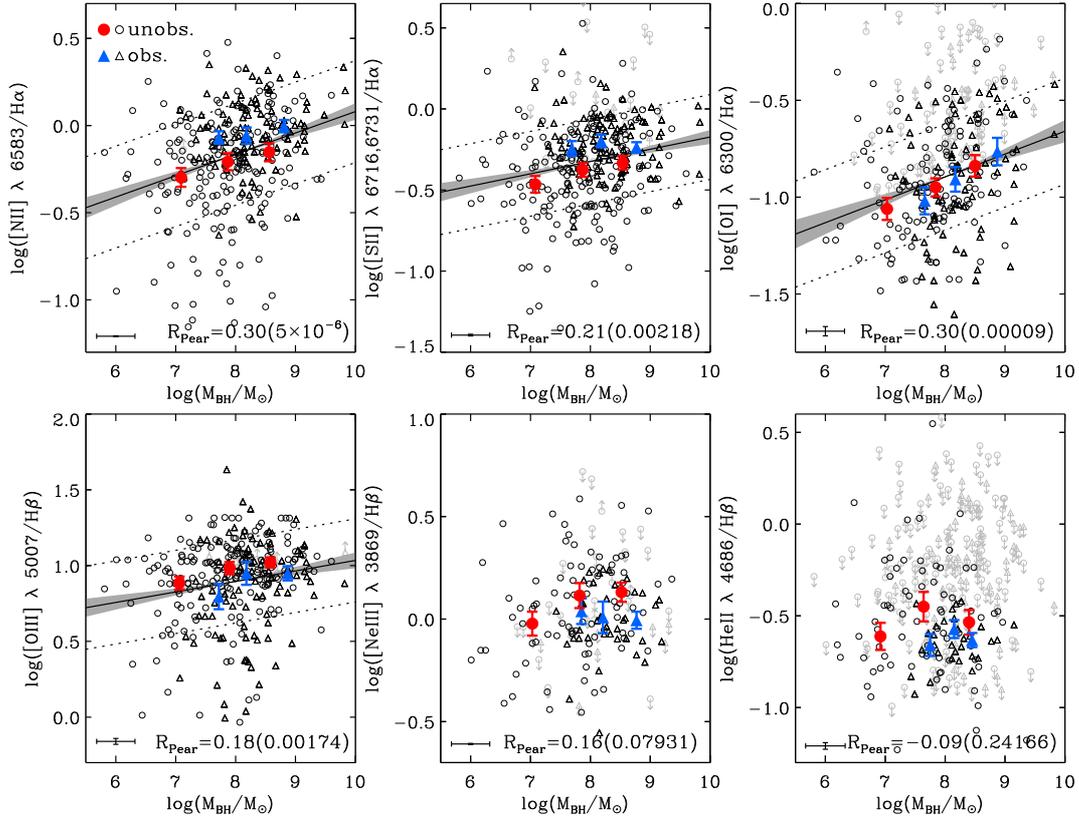}
    \caption{Optical emission line ratio versus black hole mass.  
    The format is the same as that of Fig.~\ref{fig:ratio_Edd}}
    \label{fig:ratio_Mbh}
\end{figure*}

We further study the observed anti-correlation by looking for correlations with $M_{\rm BH}$ (Fig.~\ref{fig:ratio_Mbh}). We find that the \NIIHa, \SIIHa, \OIHa, and \OIIIHb\ show positive correlations with $M_{\rm BH}$, with \textit{p}-values of $5\times10^{-6}$, 0.00218, 0.00009, and 0.00174, respectively. In order to understand whether the anti-correlation of \NIIHa\ with $\lambda_{\rm Edd}$ is stronger than the correlation of \NIIHa\ with $M_{\rm BH}$ we run the $z$-test based on the two Pearson correlation coefficients and find that the $p$-value suggests a stronger anti-correlation for the $\lambda_{\rm Edd}$ ($p$-value=0.025). We also investigate the observed anti-correlation between optical emission line ratios and $\lambda_{\rm Edd}$ with fixed $M_{\rm BH}$ ($7<log(M_{\rm BH}/M_{\odot})<8$, $8<log(M_{\rm BH}/M_{\odot})<9$). We find that \NIIHa\ is indeed significantly anti-correlated with $\lambda_{\rm Edd}$ in each of these mass bins, with $p$-value of 0.00488 ($7<log(M_{\rm BH}/M_{\odot})<8$) and $3\times10^{-7}$ ($8<log(M_{\rm BH}/M_{\odot})<9$). On the other hand, the other line ratios do not show correlation with $\lambda_{\rm Edd}$ at any fixed $M_{\rm BH}$ except \SIIHa\ which shows $p$-value of $5\times10^{-5}$ in high $M_{\rm BH}$ bin. We also test how the relationships between emission line ratios and $M_{\rm BH}$ change at fixed $\lambda_{\rm Edd}$ ($-2.5<log\lambda_{\rm Edd}<-1.5$, $-1.5<log\lambda_{\rm Edd}<-0.5$). We find that \OIHa\ and \OIIIHb\ only show weak correlation at both fixed $\lambda_{\rm Edd}$ bins with less than 1\% level of $p$-value.

\begin{figure*}
	\includegraphics[width=0.79\textwidth]{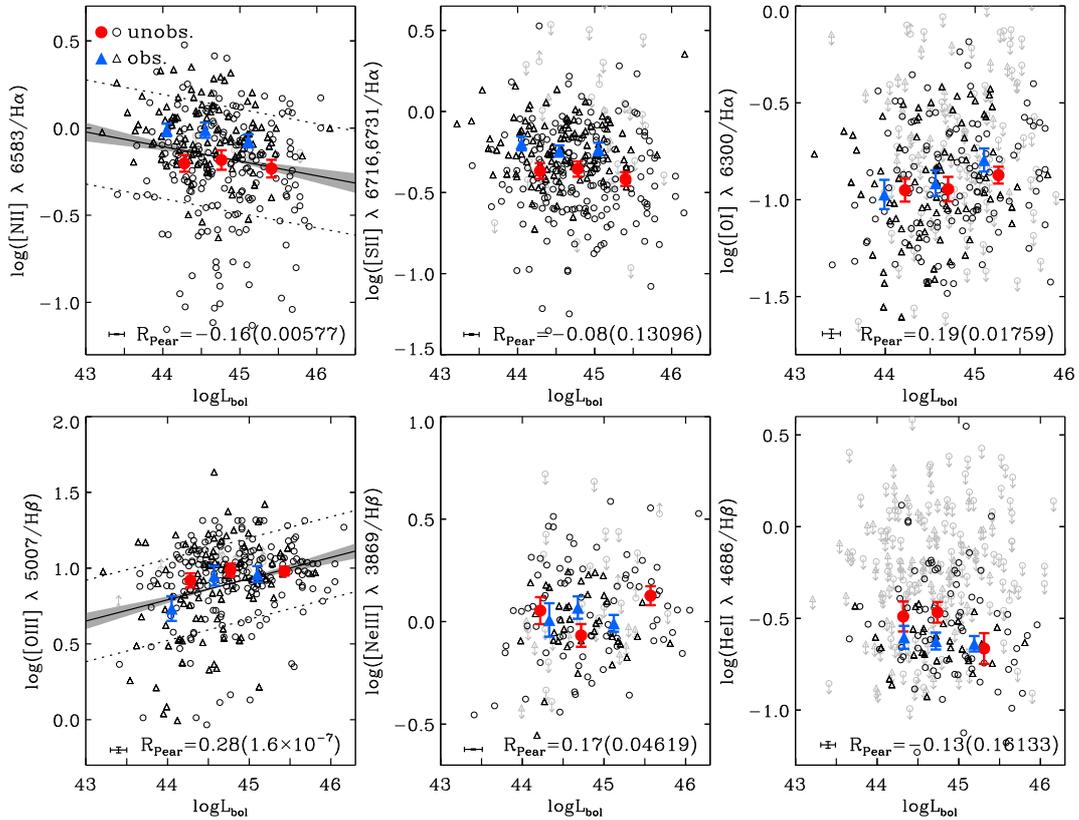}
    \caption{Optical emission line ratio versus bolometric luminosity. 
        The format is the same as that of Fig.~\ref{fig:ratio_Edd}}
    \label{fig:ratio_Lbol}
\end{figure*}

Finally, we find a negative correlation with the $L_{\rm bol}$ (Fig.~\ref{fig:ratio_Lbol}) for \NIIHa\ ($p$-value=0.00577) while a positive correlation is found for \OIIIHb\ ($p$-value=$2\times10^{-6}$). Running the $z$-test based on the two Pearson correlation coefficients for \NIIHa\ with $\lambda_{\rm Edd}$ compared to \NIIHa\ with $L_{\rm bol}$ again suggests a statistically stronger correlation ($p$-value=0.0001) in $\lambda_{\rm Edd}$. While \NIIHa\ shows correlations with $M_{\rm BH}$ and $L_{\rm bol}$, the correlation with $M_{\rm BH}$ is more significant at the less than 5\% level based on a Fisher $z$ test ($p$-value=0.036). 

In order to understand effect of the different bolometric correction, we estimate $L_{\rm bol}$ and $\lambda_{\rm Edd}$ following \citet{Marconi04} who uses bolometric correction that depends on 2-10 keV luminosity (equation 21 in their paper). The mean difference between the newly estimated $L_{\rm bol}$ and the one derived by our prescription is 0.03 dex with 0.33 dex of scatter, which gives a mean difference in $\lambda_{\rm Edd}$ of 0.03 dex (0.33 dex of scatter). We find that \NIIHa\ ($p$-value=$10^{-12}$) and \SIIHa\ ($p$-value=$1\times10^{-6}$) show significant anti-correlation with $\lambda_{\rm Edd}$.

If we adopt more steep bolometric correction curve that varies with 2-10 keV luminosity (see Figure 3 in \citealt{Marconi04}) covering wide range of bolometric correction, we may get flattened relationship in \NIIHa\ and $\lambda_{\rm Edd}$ as sources in low $\lambda_{\rm Edd}$ and high $\lambda_{\rm Edd}$ move toward each end. However, we find that the application of such extreme case of bolometric correction does not significantly change the Pearson $R$ coefficient ($-0.43$) and $p$-value ($10^{-12}$) but shows slightly moderate slope ($-0.10\pm0.01$).


\section{Discussion}

We have presented the observed relationship between the $\lambda_{\rm Edd}$ and optical emission line ratios (\NIIHa, \SIIHa, \OIHa, \OIIIHb, \NeIIIHb, and \HeIIHb) using local obscured and unobscured AGN ($ \langle z \rangle = 0.05$, $z < 0.40$) from the 70-month \textit{Swift}-BAT all-sky hard X-ray survey with follow-up optical spectroscopy. We show that there is a significant anti-correlation between \NIIHa\ emission line ratio and $\lambda_{\rm Edd}$, and this correlation is stronger than trends with $M_{\rm BH}$ or $L_{\rm bol}$ or with other line ratios. The observed trend suggests that optical emission line ratios, which are widely used to classify sources as AGN, can also be an indicator of $\lambda_{\rm Edd}$. The use of \NII\ and \Ha\ emission lines as a $\lambda_{\rm Edd}$ indicator has potential implications for high redshift obscured AGN whose $M_{\rm BH}$ and $\lambda_{\rm Edd}$ are difficult to estimate. This would require to additionally assume that any relevant physical relations that might affect our $\lambda_{\rm Edd}$ - \NIIHa\ relation (e.g., the stellar mass-metallicity, AGN outflows), do not evolve significantly with redshift. The relationship shown in this work may serve as a basis for future studies toward measuring $M_{\rm BH}$ and $\lambda_{\rm Edd}$ of individual AGN.

A number of complications arise when measuring $L_{\rm bol}$ and $M_{\rm BH}$ from a large ($N>100$) sample of galaxies. The majority of the total luminosity is emitted from the accretion disk in the extreme ultraviolet and ultraviolet energy bands \citep{Shields78, Malkan82, Mathews87}.  While we used a fixed bolometric correction from the X-ray, this correction has been observed to vary depending on $\lambda_{\rm Edd}$ \citep{Vasudevan09} and $L_{\rm bol}$ (e.g., \citealt{Just07, Green09}). This issue deserves further study, though we would expect any biases to affect all line ratios whereas we find a much stronger correlation with \NIIHa. Another complication is the use of separate methods of BH mass estimates. We note, however, that these two methods are tied to reproduce similar masses for systems where both are applicable \citep{Graham11, Woo13}, and that we find significant correlations for both type 1 and type 2 AGN, separately (Table~\ref{tab:regression}). We will explore $M_{\rm BH}$ measurements for both types of AGN via different methods in a future study.

There are several possible physical mechanisms that might lead to the trends found between $\lambda_{\rm Edd}$ and emission line ratios such as \NIIHa. \citet{Groves06} and \citet{Stern13} found a dependence of emission-line diagnostics, particularly of the \NIIHa, with host galaxy stellar mass. They postulated that this was a result of the mass metallicity relationship with more massive galaxies having more metals \citep{Lequeux79, Tremonti04, Erb06, Lee06, Ellison08, Maiolino08, Mannucci10, Laralopez10}. As more massive galaxies have more massive black holes, this follows the correlation found here with \NIIHa\ being positively correlated with $M_{\rm BH}$ and negatively correlated with $L_{\rm bol}$. 
\citet{Stern13} showed that \OIIIHb\ mildly decreases with stellar mass since reduced \OIII\ emission is expected from higher metallicity and massive systems as \OIII\ is a main coolant and the temperature will be lower in massive systems. 
The less significant correlation between \OIIIHb\ and $M_{\rm BH}$ shown in Fig.~\ref{fig:ratio_Mbh} as compared to the \NIIHa\ which scales strongly with metallicity can be explained in this context. Another interesting possibility affecting this correlation could be from higher $\lambda_{\rm Edd}$ AGN have relatively weaker \OIII\ lines, as found by the ``Eigenvector 1" relationships (e.g., \citealt{Boroson92}).  

A further possibility is that X-ray heating is inducing some of the negative correlation found between $L_{\rm bol}$ and the \NIIHa\ ratio. Ionizing ultraviolet photons produce a highly ionized zone on the illuminated face of the gas cloud while deeper in the cloud penetrating X-rays heat the gas and maintain an extended partially ionized region. Higher energy photons such as Ly$\alpha$ are destroyed by multiple scatterings ending in collisional excitation which enhances the Balmer lines \citep{Weisheit81, Krolik83, Maloney96}. Strong X-rays (i.e., harder SEDs) that heat up hot electrons in partially ionized region also enhance collisional excitation of ${\rm O}^{0}$, ${\rm N}^{+}$, and ${\rm S}^{+}$. As a result, it is expected to see high \NIIHa, \SIIHa, and \OIHa.

Alternatively, the observed anti-correlation between emission line ratios (\NIIHa\ and \SIIHa) and $\lambda_{\rm Edd}$ may be due to radiatively driven outflows in high $\lambda_{\rm Edd}$ systems. Radiatively accelerated wind is predicted to be proportional to $\lambda_{\rm Edd}$ \citep{Shlosman85, Arav94, Murray95, Hamann98, Proga00, Chelouche01}. This is consistent with the observed blueshift of broad as well as narrow absorption lines \citep{Misawa07} often seen in quasars. In the context of a prevalent outflow in high $\lambda_{\rm Edd}$ AGN, the optical-UV SED of the accretion disk is expected to be softer when $\lambda_{\rm Edd}$ is $\gtrsim 0.3$ \citep{King03, Pounds03, Reeves03, Tombesi10, Tombesi11, Slone12, Veilleux16, Woo16}.  As hot accreting gas is removed by ejecting outflows, the formation of collisionally excited emission lines is expected to be suppressed. It is important to note, however, that the anti-correlation between optical emission line ratio and $\lambda_{\rm Edd}$ is only appeared in \NIIHa\ and \SIIHa\ but not in other line ratios.

\section{Summary}
We present observed correlations between AGN Eddington ratio ($\lambda_{\rm Edd}$), black hole mass ($M_{\rm BH}$), and bolometric luminosity ($L_{\rm bol}$) and narrow emission line ratios (\NIIHa, \SIIHa, \OIHa, \OIIIHb, \NeIIIHb, and \HeIIHb) for hard X-ray selected AGN from the BASS. 
The results of this study are: \\
\begin{itemize}
	\item $\lambda_{\rm Edd}$ is anti-correlated with both the \NIIHa\ and \SIIHa\ ratios, but not with other line ratios. 
	\item \NIIHa\ exhibits a significantly stronger anti-correlation with $\lambda_{\rm Edd}$ than \SIIHa.
	\item The correlation shown in \NIIHa\ with $M_{\rm BH}$ is more significant than with $L_{\rm bol}$. 
	\item The correlation appeared in \NIIHa\ with $M_{\rm BH}$ might be a result of the mass metallicity relationship. 	
	\item The observed relationship between $\lambda_{\rm Edd}$ and \NIIHa\ ratio could be explained by considering X-ray heating processes and removal of material due to energetic outflow in the high $\lambda_{\rm Edd}$ state. 
	\item The \NIIHa\ ratio could in principle be used to measure accretion efficiencies and black hole masses of high redshift obscured AGN (equation~\ref{eq:indicator}).
\end{itemize}

\section*{Acknowledgements}
K.O. and K.S. acknowledge support from the Swiss National Science Foundation (SNSF) through Project grant 200021\textunderscore157021. M. K. acknowledges support from the SNSF through the Ambizione fellowship grant PZ00P2\textunderscore154799/1. M.K. and K. S. acknowledge support from SNFS Professorship grant PP00P2 138979/1.
C.R. acknowledges financial support from the CONICYT-Chile ``EMBIGGEN'' Anillo (grant ACT1101), FONDECYT 1141218 and Basal-CATA PFB--06/2007. 
E.T. acknowledges support from the CONICYT-Chile ``EMBIGGEN'' Anillo (grant ACT1101), FONDECYT 1160999 and Basal-CATA PFB--06/2007. The work of DS was carried out at the Jet Propulsion Laboratory, California Institute of Technology, under a contract with NASA. This research has made use of NASA's ADS Service.
 
Facilities: Swift, UH:2.2m, SDSS, KPNO:2.1m, FLWO:1.5m (FAST), Shane (Kast Double spectrograph), CTIO:1.5m, Hale, Gemini:South, Gemini:North, Radcliffe,Perkins




\bibliographystyle{mnras}
\bibliography{references.bib} 






\bsp	
\label{lastpage}
\end{document}